\documentclass[aps,preprint,amssymb,12pt,floatfix]{revtex4}
\pdfoutput=1
\usepackage{subfigure}
\usepackage{graphicx}
\usepackage{rotating} \usepackage{color}
 \usepackage{amsmath,amsthm}
\usepackage{epstopdf}
\begin{document}
\title{A structural perspective on the dynamics of kinesin motors}
\author{Changbong Hyeon}
\thanks{hyeoncb@kias.re.kr}
\affiliation{School of Computational Sciences, Korea Institute for Advanced Study, Seoul 130-722, Korea}
\author{J. N. Onuchic}
\thanks{jonuchic@ucsd.edu}
\affiliation{Center for Theoretical Biological Physics, Rice University, Houston TX, 77005-1827, USA}

% generate the title page from the info in the headers above

% 200 words max Abstract
\begin{abstract}
Despite significant fluctuation
under thermal noise, biological machines in cells perform their tasks with exquisite precision. 
Using molecular simulation of a coarse-grained model and theoretical arguments we envisaged how kinesin, a prototype of biological machines, generates force and regulates its dynamics to sustain persistent motor action. A structure based model, which can be versatile in adapting its structure to external stresses while maintaining its native fold, was employed to account for several features of kinesin dynamics along the biochemical cycle. This analysis complements our current understandings of kinesin dynamics and connections to experiments. We propose a thermodynamic cycle for kinesin that emphasizes the mechanical and regulatory role of the neck-linker and clarify issues related the motor directionality, and the difference between the external stalling force and the internal tension responsible for the head-head coordination. The comparison between the thermodynamic cycle of kinesin and macroscopic heat engines highlights the importance of structural change as the source of work production in biomolecular machines.
\end{abstract}
\maketitle

% New page
\clearpage

\section*{Introduction}
In recent years, technological advances have revolutionized our understanding to biological systems by providing unprecedented views of biomolecular dynamics at the single molecule level, bringing into reality what Feynman envisioned on nanotechnology 50 years ago \cite{Feynman61}. 
The scenery inside the cell unveiled by nanodevices displays beautiful tempo-spatial organization. 
Among a host of macromolecules that constitute the cell, of particular interest are specialized enzymes, namely biological motors that convert chemical energy stored in substrate molecules into mechanical work \cite{alberts1998cell}. 
From transport motors that move along cytoskeletal filaments (kinesin, dynein, myosin) \cite{Hirokawa98Science,vale2003Cell} to the sophisticated protein production machinery (ribosome) \cite{YonathCell01,NissenScience00}, biological motors play vital roles in controlling cellular functions, such as gene expression, intra and inter-cellular trafficking, cell motility, and mitosis \cite{AlbertsBook}.
A malfunction in these motors can cause detrimental effects in the cell.   
In the highly dissipative nanoscale environment that immediately randomizes any ballistic motion, 
it is amazing to see how these evolutionally tailored molecules carry out biological functions with exquisite precision. 
Elucidating physical principles that bring molecular machines into action is one of the most challenging problems in modern biology. 

Despite notable progress made in current nanotechnology, a precise vision on biological motors still remains elusive due to ineluctable thermal noise in nanoscopic measurements.  
The questions begged from experiments demand additional details of the structures and dynamics of molecules \cite{Alberts1992Cell,Block07BJ}, which require greater spatial, temporal resolutions and better control than current techniques can provide \cite{Greenleaf07ARBBS}.
To this end, simulations using molecular models of biological motors \cite{Koga06PNAS,Hyeon06PNAS,Hyeon07PNAS,Hyeon07PNAS2,Chen10PNAS,Tehver10Structure,Takano10PNAS,kravats11PNAS} can supplement the current experimental findings and make insightful predictions amenable to future experimental investigations.  

In the study of biological motors, however, computational approach using molecular simulations has been relatively rare compared with other area focusing on folding dynamics of small sized proteins. 
This is partly because the typical size and time scale associated with biological motors are far greater than usual computational approaches have dealt with. 
Nevertheless, biomolecular dynamics have a hierarchical structure in terms of their time scale, characterized with a large time scale separation between atom ($\sim fs$), residues ($\sim ps$) and domain motions ($\gtrsim \mu s- ms$).    
To study the dynamics associated with a particular range of time scale, faster degrees of freedom can be renormalized into an effective degree of freedom. 
As long as the topological feature characterizing molecular structures are unchanged, coarse-graining the atomistic details does not essentially alter the global motion responsible for biological function.
Thus, the biology can employ the strategy of condensed matter physics that simplifies the phenomenon of interest based on its time scale and relevant degrees of freedom \cite{Anderson:1997,hyeon2011NatureComm} can be applied to the biology.  
Together with the current technical advances in single molecule measurements on diverse biological motor systems, 
dynamics generated from coarse-grained representation of motors allow one to grasp the gist of design principle under which each biological motor performs its biological function. 
In this perspective, recent efforts made in studying the dynamics of various biological motors using coarse-grained models \cite{Koga06PNAS,Hyeon06PNAS,Hyeon07PNAS,Hyeon07PNAS2,Takano10PNAS,Tehver10Structure,Chen10PNAS,kravats11PNAS} are quite promising. 

Physical conditions that govern the dynamics of objects in the nanoscopic to microscopic world are fundamentally different from those for the macroscopic objects (see SI text and Fig.~S1).  
To this end, the conceptual framework developed from the theory of protein folding \cite{OnuchicCOSB04,Shakhnovich06ChemRev,Dill08ARB,Thirumalai10ARB} and polymer physics \cite{deGennesbook}, 
which handle the complex dynamics of chain molecules over multi-scale, is of great use to decipher the design principles of biological motors. 
Here, we will discuss the dynamics of kinesin by comparing our insight gained from structure-based coarse grained model of kinesin with the experimental findings in each stage of biochemical cycle.   
Lastly, we will construct a thermodynamic cycle of kinesin using the neck-linker, a key mechanical element responsible for work generation, and highlight the importance of structural change in producing net mechanical work per cycle and in the regulation.

\section*{General overview: biochemical states of kinesin}
Kinesin-1 (hereafter kinesin) transports cellular organelles along the network of cytoskeletal filaments and play a central role in spatially organizing the cellular environment.
Two identical motor domains, linked via coiled-coil stalk, alternately exploit the free energy generated from ATP molecules to produce characteristic 8-nm steps.
Single molecule experiments have revealed that at the physiological condition, kinesins can processively travel about 100 steps in a ``hand-over-hand" fashion at a velocity of $v\approx 1$ $\mu m/s$ \cite{Block03Science,Nishiyama01NCB,Yildiz04Science}.

Recently, a growing number of kinesin structures, with various ligand states, has shed light on more detailed mechanisms of how kinesins function.
A series of crystal structures suggest how kinesins adapt their conformation to varying nucleotide (NT) states via $\rightarrow$[$\phi$$\rightarrow$ATP$\rightarrow$ADP$\cdot$Pi$\rightarrow$ADP]$\rightarrow$ (Fig.~S2). 
It is currently surmised that the local conformational changes due to the NT chemistry are  amplified via the switch II helix ($\alpha4$) to regulate (i) the neck-linker (NL) conformation and (ii) the binding affinity to MTs  \cite{Sindelar10PNAS}. 
For kinesin-1, the states of NL and MT affinity are the two major components that determine the logic of biochemical cycle.   
Depending on the NT state, the NL is in either ordered (zippered) or disordered (unzippered) state; the binding affinity of kinesin head domain to MTs is either strong or weak. 
The relationship between these two \emph{binary switches} in a single head, summarized in the Table~\ref{Table} (see also Figs. ~S2b and S2c), forms a basic logic for the kinesin dynamics.

\begin{figure}
\begin{center}
\includegraphics[width=5.0in]{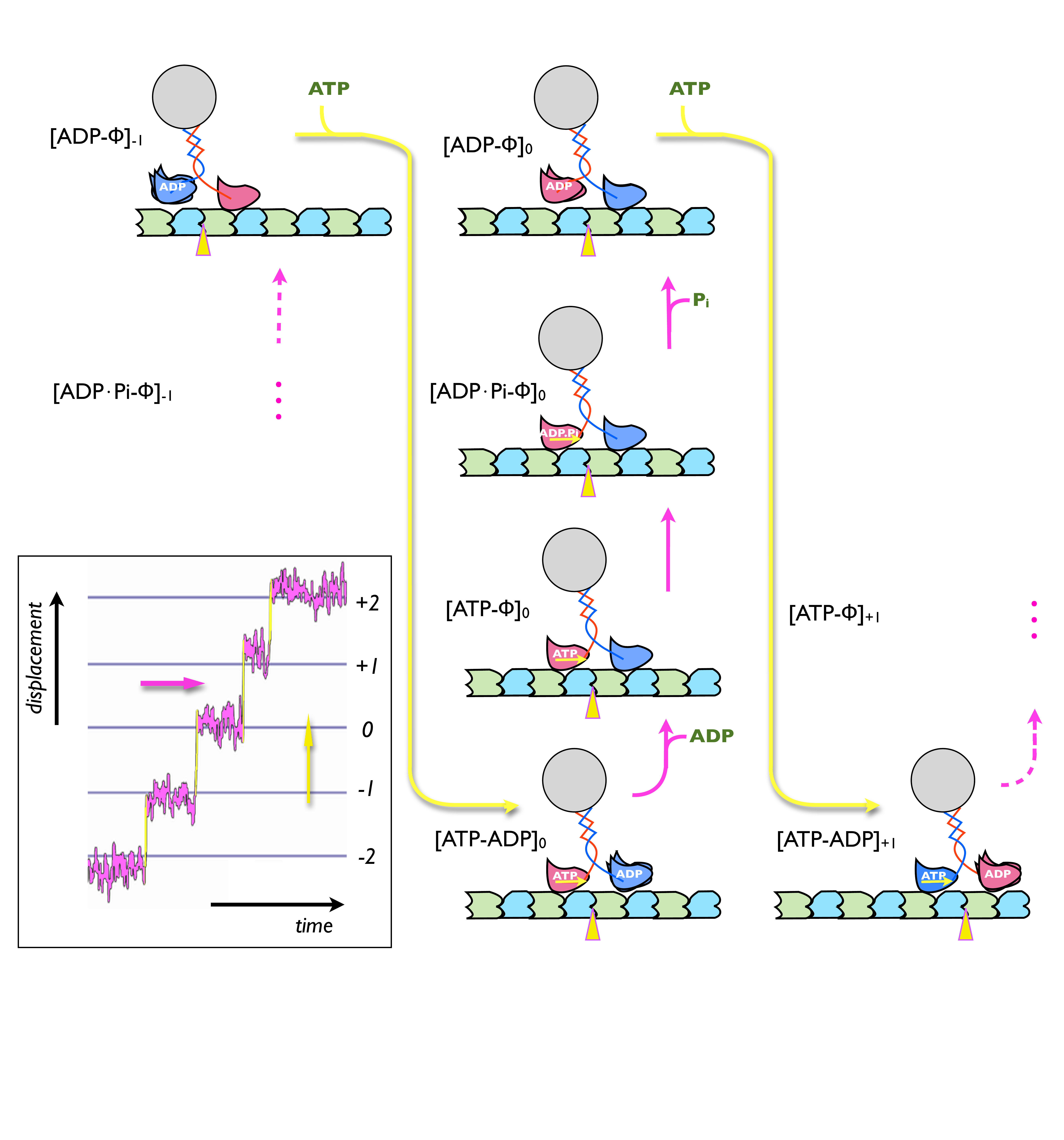}
 \caption{Biochemical cycle of a kinesin motor. 
The state of kinesin changes along the cycle. 
The subscript $i$ in the notation [NT-NT]$_i$ denotes the position on MT to which the kinesin binds.  
In the diagram the internal conformational states that do not alter the kinesin position are aligned vertically with purple arrows, and the step is depicted with yellow arrows. 
The inset shows a time trace from laser optical tweezers measurement. }
 \label{cycle_kinesin}
 \end{center} 
\end{figure}

The four different NT states ($\phi$, ATP, ADP$\cdot$Pi, ADP) at the catalytic site enable one to assign, in principle, 16 different states to the kinesin dimer; however, most of the possible dimer states has to be excluded. 
First, the ADP-ADP state (the notation (NT)$_L$-(NT)$_R$ represents the NT state of trailing and leading head from left to right and we assume kinesins move from left ($-$ end) to right ($+$ end). Below we omit $(\cdots)_L-(\cdots)_R$ in the notations), corresponding to the ``weak-weak" MT binding state, should be excluded since at least one of the heads should hold MTs tightly to remain on the MT track.   
Second, due to the topological constraint imposed by the NL zipper in the leading head (Fig.~S2c), the NL of the leading head cannot be in an ordered state, which prohibits the $\phi$-ATP, ATP-ATP, and ADP-ATP (or $\phi$-ADP$\cdot$Pi, ATP-ADP$\cdot$Pi, and ADP-ADP$\cdot$Pi) states from the possible dimer states.  
$\phi$-ADP and $\phi$-$\phi$ states are also unlikely to exist because the time spent for ATP to dissociate from the trailing head (ATP$\rightarrow$ADP$\cdot$Pi$\rightarrow$ADP$\rightarrow\phi$) ($>1$ sec \cite{Ma97JBC}) is longer than the timescale associated with ADP dissociation from the leading head ($\lesssim 10$ ms \cite{Ma97JBC}).  
Therefore, only four states (ATP-ADP, ATP-$\phi$, ADP$\cdot$Pi-$\phi$ and ADP-$\phi$)  are permissible out of the 16 states. 
%A scenario that ATP hydrolysis at the trailing head occurs before the ADP release at the leading head, which produces ADP$\cdot$Pi-ADP state, is excluded by the report that the release of ADP from the leading head occurs rapidly ($\approx 300$ $s^{-1}$ \cite{Moyer98BC}) followed by binding of ATP to its partner head. 
The conventionally accepted kinesin cycle is schematized in Fig.\ref{cycle_kinesin} \cite{Cross00PTRSL,Xing00JBC,BlockPNAS06}.

To sustain a constant material flux along the biochemical cycle, kinesins need continuous supply of ATP (source) and removal of ADP and Pi (sink) (see SI text for the discussion on nonequilibrium steady state thermodynamics). 
In the kinesin cycle depicted in Fig.~\ref{cycle_kinesin}, ATP binding induced stepping ([ADP-$\phi$]$_{i-1}$$\rightarrow$[ATP-ADP]$_i$) is the only mechanical motion (yellow arrows in Fig.~\ref{cycle_kinesin}) that can be detected with SM measurements (the inset of Fig.~\ref{cycle_kinesin}). 
All other steps (purple arrows in Fig.~\ref{cycle_kinesin}) are associated with internal chemistry. 
In contrast to the common notion for macroscopic heat engines where heat released from combustion is directly used to expand the volume of cylinder and provide the power stroke (see Fig.~S1 and SI Text), the heat released from the ATP hydrolysis step in the catalytic site itself is not the main driving force of the stepping motion.  
A series of structures that vary with NT state and biochemical data indicate that the ATP hydrolysis itself alters neither the NL state nor the affinity to MT. 
For kinesins, the free energy released from the ATP hydrolysis at the catalytic site is estimated to be relatively small $\approx 4$ $k_BT$ \cite{Gilbert94Biochem,Hackney05PNAS,Hyeon09PCCP}. 
Even though the free energy change due to ATP hydrolysis in the catalytic site is small, the irreversibility of the enzymatic cycle set by the ATP hydrolysis has an important implication for the robust action of biological motors \cite{Hopfield74PNAS}.

\section*{ATP binding induced conformational change} 
ATP binding to a catalytic site creates new contacts to the residues that constitute the catalytic site ($\Delta H<0$), and reduces the local fluctuations ($\Delta S<0$). 
As long as the stabilization free energy ($\Delta G=\Delta H-T\Delta S$) remains negative, the ATP binding is favored. 
Previously, using the molecular simulation we have shown for the catalytic domain of protein kinase A that local compaction of ligand binding pocket upon ATP binding induces a global open-to-closed transition \cite{Hyeon09PNAS}. For kinesin, despite the presence of a number of high resolution cryo-EM or crystal structure data \cite{parke2010JBC,Sindelar10PNAS}, it is still an open question  how explicitly the intramolecular signal transduction leads to the disorder-order transition in the NL. 
Nevertheless, the difference in the contact maps before and after the formation of NL zipper (Fig.~\ref{Stepping}a) can be used to calculate the folding landscape of kinesin for each case.
From the perspective of energy landscape, visualized in terms of the centroid of tethered head whose motion is constrained by the NL (see Fig.~\ref{Stepping}b) \cite{Hyeon07PNAS2}, it could be argued that ATP binding reshapes the energy landscape from the disordered state ($F_{disorder}(\{\vec{r}\})$) to the ordered state ($F_{order}(\{\vec{r}\})$), and gives rise to a conformational force ($\vec{f}_{conf}(\{\vec{r}\})=-\vec{\nabla}F(\{\vec{r}\})$), which can be interpreted as the \emph{power stroke}.  
The NL zipper formed in the MT-bound head biases the diffusive motion of the tethered head toward the (+)-end direction, which has also been termed the NL docking model \cite{ValeNature99}.

One well-known objection to the NL docking model as a driving force for the kinesin step is that the free energy gain calculated from EPR measurement $\Delta G_{dock}\approx -1.2$ $k_BT$ \cite{Sindelar02NSB,Rice03BJ,Block07BJ}, is energetically insufficient to drive the kinesin step. 
However, as will be discussed in the next section, the energetic contribution of the NL docking (or power stroke) to the stepping motion needs not necessarily be large as long as the NL zippered state can provide enough anisotropic bias, so that the tethered head can reach the next MT binding site through the diffusive search.
Interestingly, recent SM experiments and computation have suggested that a structural motif called cover-neck bundle (see Fig.~\ref{Stepping}a) can guide the early stage of the NL docking dynamics \cite{Hwang08Structure,khalil2008kinesin}.  
%$\Delta G_{dock}$ value estimated from the EPR experiment should be taken as a lower bound of the free energy change since the measurement was made on an EPR labeled kinesin monomer on MTs. 
%Given that tethered head dangling at the other end of NL can produce greater configurational states, much larger entropy reduction upon docking ($T\Delta S_{dock}$) is expected for a kinesin dimer while the enthalpy change ($\Delta H_{dock}$) remains the same. 
In addition, a simple back-of-envelope calculation using available thermodynamic/kinetic data provides a much greater estimate of the net stability difference involving the step from [ADP-$\phi$]$_{i-1}$ to [ATP-ADP]$_i$ as $\Delta G_{step}\approx (-10\sim -12)$ $k_BT$ \cite{Hackney05PNAS,Hyeon09PCCP,Derenyi11BJ} for quasi-static processes, which agrees with the work required to stall the kinesin motion $W=(6-7)$ pN $\times $8 nm $\approx 12$ $k_BT$. $\Delta G_{step}$ is expected to consist of $\Delta G_{dock}$ and other free energy contributions due to the interaction between the ADP containing tethered kinesin head and MT surface.
In fact, how tightly or loosely the free energy of molecular fuel is coupled to the functional motion is the key question that has been asked for many different motors \cite{oosawa2000loose}. Similarly, contrasting the mechanism of power stroke (or conformational change) with Brownian ratchet (or diffusive motion) may be a useful way to dissect the role of energetics and flexibility of molecular machines in performing their functional motion.

\section*{Stepping dynamics}
The stepping motion in search of the next MT binding site is the most dramatic and functionally important step in the 
kinesin cycle. 
Nevertheless, because the time scale associated with this motion ($\sim (10-100)$ $\mu sec$) is transient compared with that of the entire cycle ($>10$ $ms$), details of the dynamics has been elusive and under intensive debate \cite{Coppin96PNAS,Nishiyama01NCB,Cross05Nature,Block07BJ}. 
The simple looking jump from one binding site to another, in fact, contains many complicated dynamics at molecular level: 
(i) The NL restrains the dynamics of the tethered head.
(ii) Depending on the NT state, the NL conformation is either in disordered or ordered state, which influences the range of space for the tethered head to explore.   
The disorder-order transition of the NL in the MT-bound head can rectify the diffusive motion of tethered head towards the next MT binding site.  
(iii) MTs provide multiple binding sites for kinesins. 
Given the microscopic constraints and dynamics described above, what is the dynamical or structural origin of substeps observed in some of the experiments \cite{Coppin96PNAS, Nishiyama01NCB}? To what portion of entire stepping motion is contributed by the motion due to NL docking and diffusive search? 

\begin{figure}
\begin{center}
\includegraphics[width=4.0in]{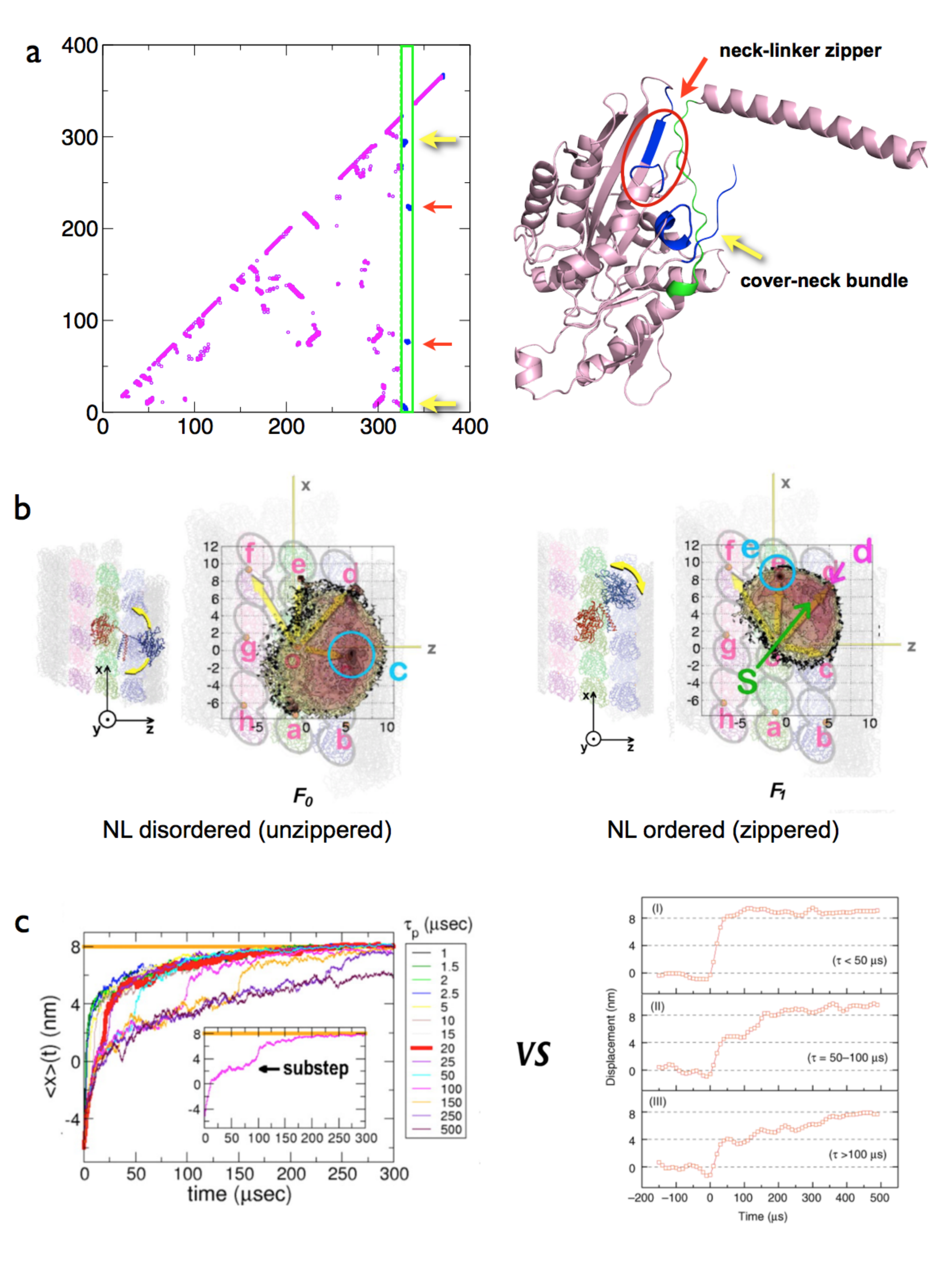}
 \caption{Stepping dynamics of kinesin. 
{\bf a}. Difference of contact maps calculated with ordered and disordered NL reveals the contacts involving the NL docking and cover-neck bundle. 
Red (solid) arrows indicate the NL zipper contacts on the contact map and in the kinesin structure. 
Yellow (shady) arrows show the location of cover-neck bundle, where the N-terminal part of NL is sandwiched between two chain segments (residues 2-8 and 289-296). 
{\bf b}. PMF of the tethered head on the MT surface computed for disordered (left) and ordered NL (right) of the MT-bound head.      
{\bf c}. The average time traces from the Brownian dynamics simulation of a quasi-particle with varying $\tau_P$ (the panel on the left) are compared with the average time traces measured using optical tweezers by Yanagida and coworkers (the three panels on the right) \cite{Nishiyama01NCB}. In Ref. \cite{Nishiyama01NCB} the individual SM time traces were partitioned into three groups depending on the stepping time ((i) $t<50$ $\mu s$ , (ii) 50 $\mu s$ $<t<$ 100 $\mu s$ , (iii) $t>100$ $\mu s$) and averaged in each group. The panels on the right are adapted from the reference \cite{Nishiyama01NCB}. }
 \label{Stepping}
 \end{center}
\end{figure}

Probing the stepping dynamics at molecular detail is even now an extremely demanding task because of the limited spatial and temporal resolutions of the current instrumentation \cite{Coppin96PNAS,Nishiyama01NCB,Cross05Nature,Block07BJ}. 
Therefore, a molecular model is of great help to study this particular problem \cite{Hyeon07PNAS2}. 
Kinesin models of ordered and disordered NL, created by adjusting the strength of the NL native contacts (Fig.~\ref{Stepping}a), were used to calculate the potentials of mean force (PMFs) and to visualize how the formation of NL zipper changes the search space for the tethered head. 
Under the disordered NL, the tethered head spans the broad area around the side-way binding site ($c$ in Fig.~\ref{Stepping}b), but cannot reach the next target binding site ($e$ in Fig.~\ref{Stepping}b). 
When the NL is in the docked state, the search space is confined preferentially to the forward direction relative to the MT-bound head, guiding the tethered head to the target binding site. 
A theoretical consideration based on two competing time scales can be proposed; if the disorder-order NL docking transition is slower than the time scale associated with exploring the MT surface, so that the tethered head can fully explore the MT surface while the NL is still in the disordered state, then the side-way binding site ($c$ in Fig.~\ref{Stepping}-b) can transiently trap the swinging tethered head; and consequently a substep emerges in the averaged time trace. 
This scenario can be simulated using the dynamics of a quasi-particle on a time dependent free energy surface $F(\vec{r},t)/k_BT=-\log{\left[t/\tau_p\cdot e^{-F_0(\vec{r})/k_BT}+(1-t/\tau_p)\cdot e^{-F_1(\vec{r})/k_BT}\right]}$ for $t\leq \tau_p$ and $F(\vec{r},t)=F_1(\vec{r})$ for $t> \tau_p$, which combines the PMFs at the two extreme cases; $F_0(\vec{r})$ under disordered NL and $F_1(\vec{r})$ under ordered NL. 
Here, $\tau_p$ is a parameter for the duration of NL docking transition. 
For a quasi-particle representing the centroid of tethered head with a diffusion constant $D=0.2$ $\mu m^2/s$, the averaged time traces exhibit the signature of a substep when $\tau_p$ is greater than $\approx$ 20 $\mu s$ (Fig.~\ref{Stepping}c left), which is similar to the one observed by Yanagida and coworkers \cite{Nishiyama01NCB} (see Fig.~\ref{Stepping}c right). 
Our theoretical study using multiscale simulation method suggests that
the ratio between the time scale of NL zipper dynamics and the exploration time on MT surface determines the detailed pattern of the stepping \cite{Hyeon07PNAS2}. 

Interestingly, the recent experiment by Yildiz \emph{et al.} for wild type kinesin using a quantum dot labeling \cite{Yildiz08Cell} displayed $\approx$ 13\% of 6 nm displacement, which is likely to be the signature of sideway steps to the neighboring protofilament. 
Given the longstanding debate on the existence of kinesin substep \cite{Coppin96PNAS,Nishiyama01NCB,Cross05Nature,Block07BJ}, 
re-investigating kinesin time traces using an instrument with improved temporal and spatial resolution would be of great interst.

\section*{Factors determining motor directionality}
Despite a remarkable similarity among the head domain structures from $\approx $ 50 different kinesins belonging to the same family, the functions of kinesins are diverse; kinesin-1 moves towards $(+)$-end, kinesin-14 (C-terminal motor) moves towards $(-)$-end, kinesin-3 (KIF1A) undergoes diffusive motion along MTs \cite{HirokawaSCI04}, and kinesin-13 quickly diffuses along MTs to reach the ends to depolymerize tubulins \cite{helenius2006Nature}.  
Given the structures of a motor with varying NT states and motor track, what determines the directionality of the motor?  
Brownian ratchet models suggests three basic conditions for the unidirectional motion: (i) asymmetric potential, (ii) thermal fluctuation, and (iii) athermal fluctuation (flashing ratchet or fluctuating potential) \cite{Astumian97Science}. 

In kinesins, these three conditions are realized through the interplay of several factors. 
First, the interaction of kinesins in ADP-ADP state with MTs results in dissociation of an ADP from one of the two heads, breaking the symmetry of NT state of the two heads.
Throughout the entire cycle, NT states remain asymmetric in the two motor heads.   
Second, kinesin motor domain binds MT in preferred direction, so that the NL in ordered state points toward the (+)-end of MTs. 
Third, as clearly demonstrated in Fig.~\ref{Stepping}b, the ordered state of NL restrains the search space of swinging head toward the (+)-end direction. 
Finally, the irreversibility of ATP hydrolysis sets the arrow of time, breaking the time reversal symmetry. 

Given the thermal and athermal (ATP-cycle) fluctuations that are provided as basic driving force for the conformational change in biological enzymes, 
the directionality of a biological motor is determined by the structure and conformational dynamics and its specific interaction with cytoskeletal filaments and other proteins \cite{Roostalu11Science}.

\section*{Conformational flexibility$-$ A strategy to avoid high free energy barriers} 
A stark difference in the dynamics of biological motors from that of macroscopic machines is that the intrinsic energy scale responsible for the biological assembly is comparable to the thermal energy (see also SI Text).
It is well entrenched that free energy gaps between functional states of biomolecules are marginal \cite{Toozebook}, thus a biomolecule can easily adapt its conformation to environmental changes due to chemical transformation of molecular fuels and external tension.  
In addition, the flexibility of biomolecules makes their characteristic dynamics entirely different from that of macroscopic system. 
For instance, binding between two partner molecules such as enzyme and substrate protein does not necessitate an absolute geometrical complementarity \cite{Fischer1894}. 
As suggested by Koshland, it is more likely for flexible biomolecules that interactions between biomolecules are realized through a mutual adaptation of the structures \cite{Koshland58PNAS}. 

\begin{figure}
\begin{center}
\includegraphics[width=4.50in]{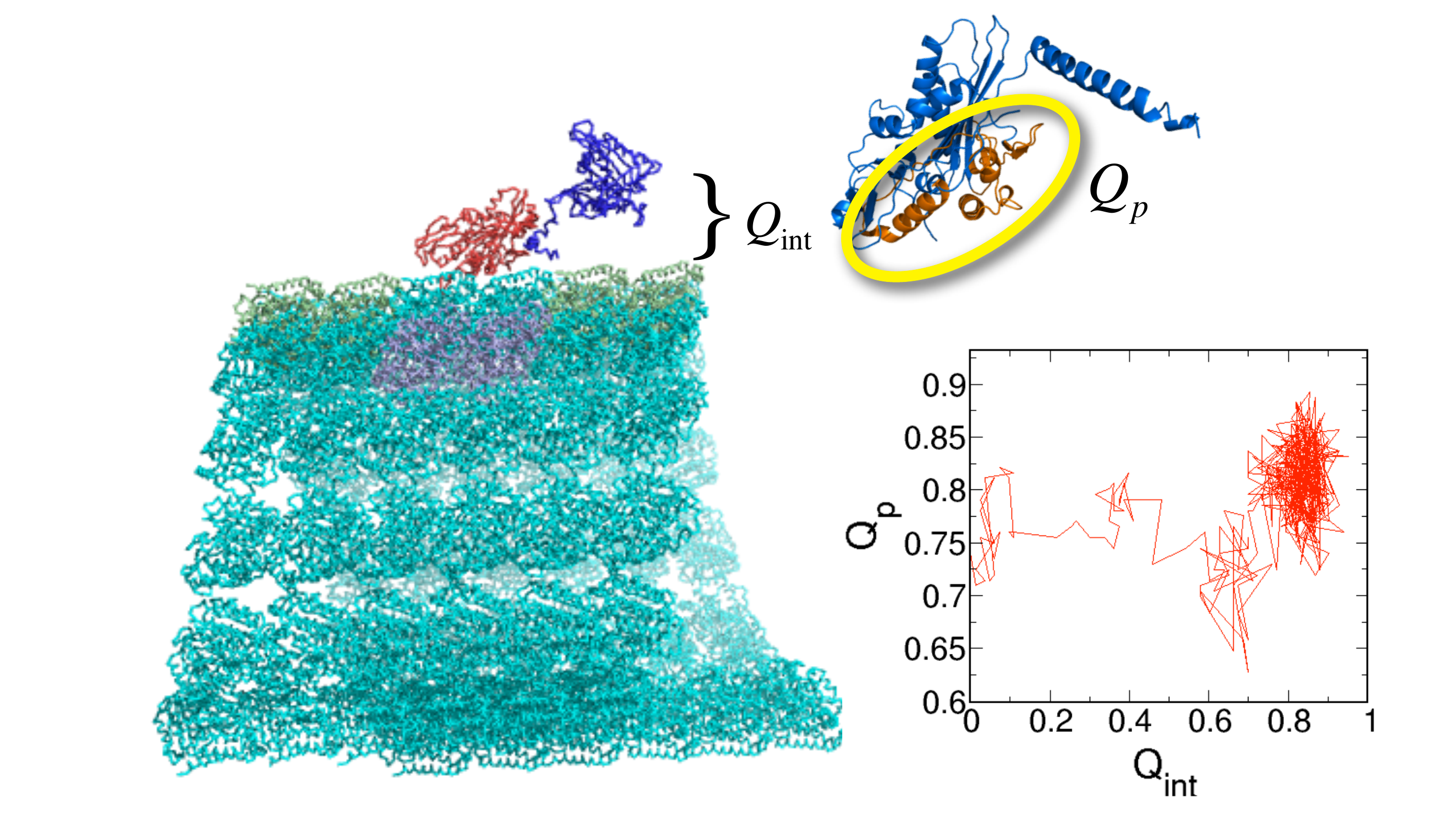}
 \caption{Facilitation of kinesin binding to MTs through partial unfolding and refolding of MT-binding motifs.
Binding process is monitored using the fraction of native contacts within the MT-binding motifs made of $\alpha 4$, $\alpha 5$, $\alpha 6$, $L12$ and $\beta 5$ ($Q_p$) and the fraction of interfacial contacts between kinesin and MT binding site ($Q_{int}$). 
The inset shows an exemplary trajectory exhibiting partial local unfolding and refolding of MT-binding motifs prior to the binding.}
 \label{binding}
\end{center}
\end{figure}

The mechanisms of partial unfolding/refolding and fly-casting to avoid large binding free energy barrier have been theoretically postulated \cite{Shoemaker00PNAS,Miyashita03PNAS,levy2005JMB} and experimentally implicated  \cite{Spolar94Science,Dyson05NRMCB} in binding processes associated with protein-protein or DNA-protein interaction, and in proteins with intrinsically disordered region \cite{dunker2002BC}. Our calculation suggests that kinesins also adopt this strategy to march along the MT by efficiently identifying  the next MT binding site. 
Although the transient disruption of local structure may be difficult to detect, such dynamics could be important if free energy barrier associated with protein-protein recognition is too high. 
Indeed, the simulation of binding process of our kinesin model on MTs showed that the partial disruption of the internal structure facilitates the binding dynamics of kinesin \cite{Hyeon07PNAS2}. 
The binding events monitored by using the native contacts of the kinesin's MT-binding motifs ($Q_p$) and the interface contacts between the kinesin and MT ($Q_{int}$) show transient decrease of $Q_p$ value prior to binding (Fig.~\ref{binding}). 
The transient partial local unfolding and refolding help the kinesin bind the MTs by bypassing otherwise a high free energy barrier.

\section*{Out-of-phase head-head coordination in the enzymatic cycle.}
While biomolecules generally fold to function, interaction with other molecules can restrain a part of molecular structure, deliberately suppressing its ability to fold. 
Such a strategy is found in the design principles of biological motors that consist of multiple domains. 
Especially, dimeric kinesin maintains the asymmetrical chemical state in two motor heads and accomplish processive steps along MTs. 

\begin{figure}
\begin{center}
\includegraphics[width=3.8in]{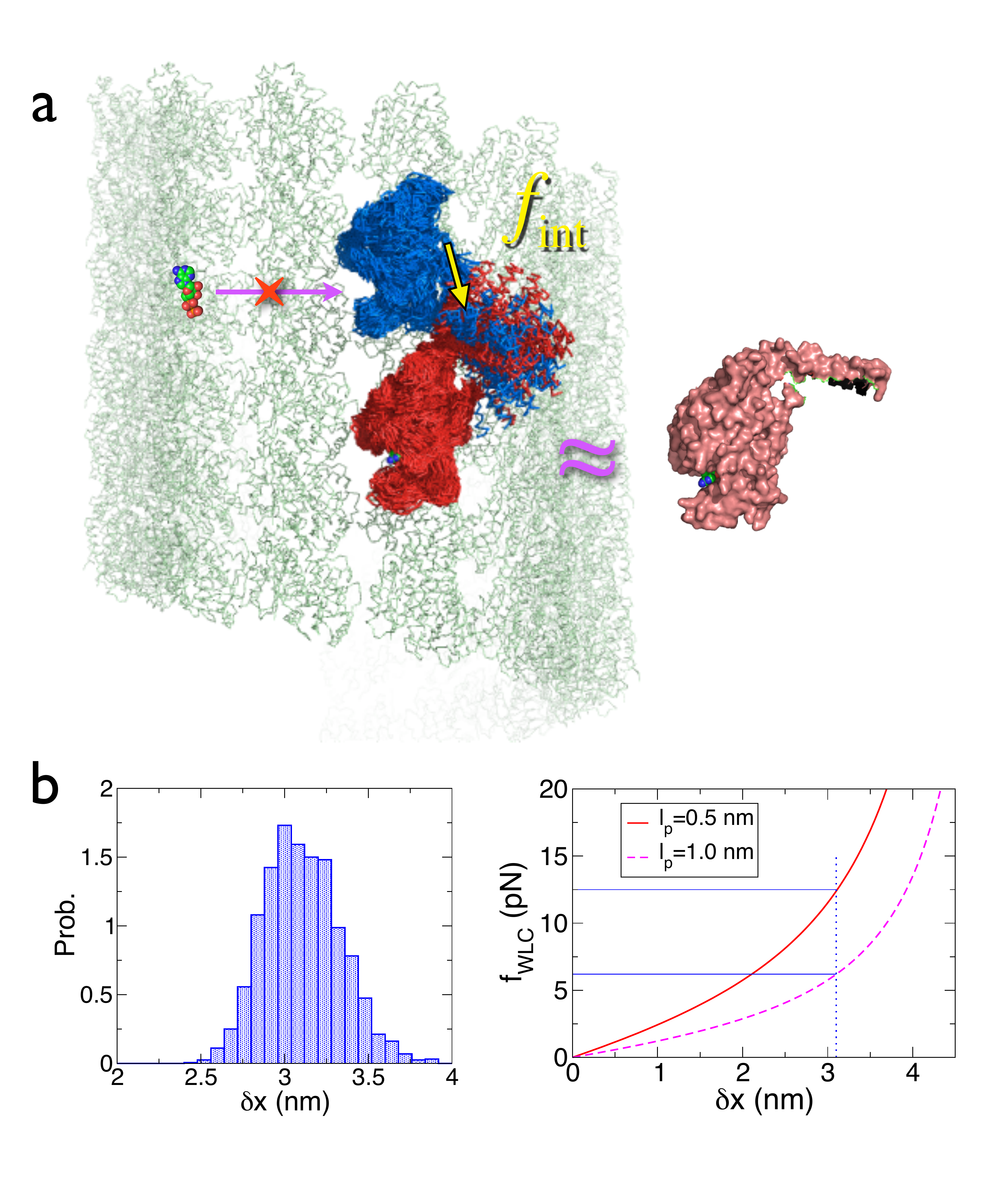}
 \caption{Internal tension regulated inhibition of ATP binding to the leading kinesin head. 
{\bf a.} For the two head-bound kinesin, the internal tension ($f_{int}$) built on the NL of the leading head disrupts the catalytic site and inhibits the premature binding of ATP, whereas the trailing head configuration is close to the native state, which is shown on the right for comparison. C$_{\alpha}$-RMS deviation of head domain excluding $\alpha 6$ helix (residues 2-315) is 1.8 \AA\ for trailing head and 3.8 \AA\ for leading head.  
{\bf b.} $f_{int}$ value can be estimated by using force-extension relation of worm-like chain model. The extension of the NL in the leading head ($\delta x$) is $\approx 3.1\pm 0.8$ nm (distribution on the left panel). When $L_c=15$ aa$\times 0.38$ nm/aa$\approx 5.7$ nm and $l_p=0.5-1.0$ nm with $\delta x=3.1$ nm are used, one can estimate the internal tension, $f_{int}\approx 7.5-12.5$ pN (right panel). }
 \label{internaltension} 
\end{center}
\end{figure}

For kinesins to maintain a high processivity, the enzymatic cycle of each head is kept out-of-phase. 
A quest to understanding the molecular origin of the head-head coordination has been pursued over the last decade. 
It has been surmised based on a series of experiments that the tension on NL is responsible for the out-of-phase coordination between the two motor domains. 
Nevertheless, it was not straightforwardly determined whether such coordination is realized due to the facilitated detachment of trailing head induced by forward tension \cite{Hancock99PNAS} or due to the rearward tension induced inhibition of ATP binding to the leading head \cite{Uemura03NSB,BlockPNAS06}.  
Remarkably, treating the MT surface as a geometrical constraint on which the kinesin dynamics occurs, 
our molecular simulation using $C_{\alpha}$-based coarse-grained model of kinesin suggests that, when both heads of kinesin are bound to the MT binding site as in ATP-$\phi$ or ADP$\cdot$Pi-$\phi$ states, 
the NL of the leading head that is stretched backward perturbs the catalytic site of the leading head away from its native-like configuration, whereas the catalytic site of trailing head remains intact \cite{Hyeon07PNAS}. 
The internal tension ($f_{int}$) exerted through the NL, estimated using the extension of the NL calculated from the simulation ($\delta x$) and the force-extension relationship of wormlike chain model ($f_{int}=k_BT/l_p\times [1/4(1-\delta x/L)^2-1/4+\delta x/L]$), is $8-15$ pN depending on the $l_p$ value assigned \cite{Hyeon07PNAS}.  
The structure-function relationship implies that the deformation of catalytic site results in the loss of ATP binding affinity to the catalytic site, thus inhibiting the premature of binding of ATP.   
Such an ATP inhibition state is maintained as long as the two heads remain bound to MTs. 
The deformed leading head catalytic site is restored only after the $\text{P}_\text{i}$ is released to change the trailing kinesin head from strong to weak affinity state with respect to MTs.  
Besides the inhibition of ATP binding, the disruption of the catalytic site in the leading head regulates the off-rate of ADP ($k_{off}^{ADP}$), which was also discussed by Uemura \emph{et al.} \cite{Uemura03NSB}. 
Note that $k_{off}^{ADP}=75-100$ $s^{-1}$ \cite{Ma97JBC}, 300 $s^{-1}$ \cite{Moyer98BC} in the leading head but $k_{off}^{ADP}=1$ $s^{-1}$ in the trailing head.  
The asymmetric strain induced regulation mechanism \cite{BlockPNAS06,Uemura03NSB}, the molecular origin of the head-head coordination, and processivity unique to the kinesin-1, is the consequence of the interplay between several topological constraints imposed on kinesin/MT complex structure. 
By explicitly showing the perturbed configuration at the catalytic site of the leading head (Fig.~\ref{internaltension}), our computational study using the simple structure based model supports the experimental proposal of the rearward strain regulation mechanism between the two motor heads \cite{Uemura03NSB,Hyeon07PNAS,Yildiz08Cell,Miyazono10EMBOJ, BlockPNAS06}.

\section*{Diagram of internal tension versus extension curve for the neck-linker.}
\begin{figure}
\begin{center}
 \includegraphics[width=5.0in]{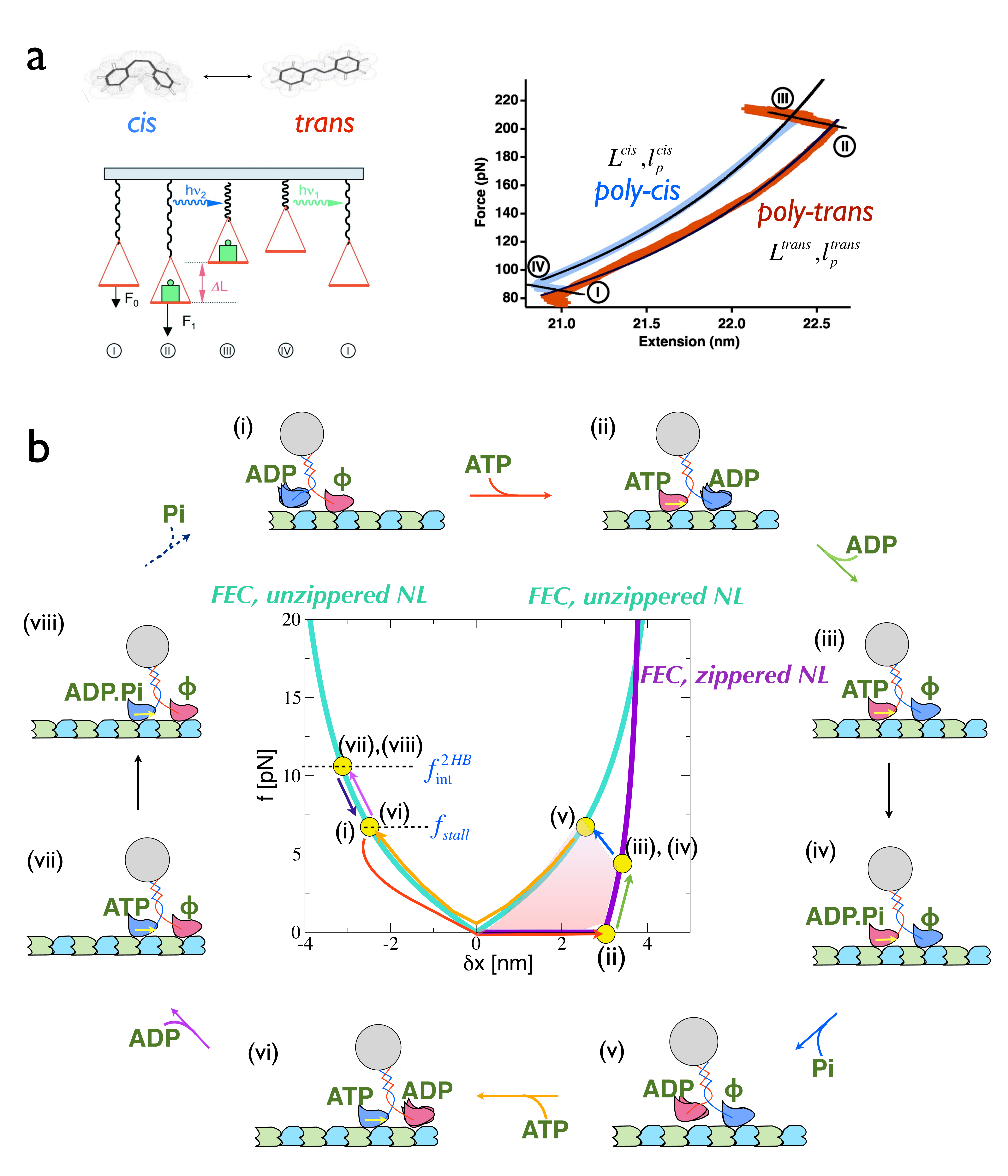}
 \caption{{\bf a.} A molecular motor devised by using optomechanical cycle of polyazopeptide that undergoes $cis\leftrightarrow trans$ transition upon photon absorption/emission. The figure is adapted from \cite{Hugel02Science}.    
{\bf b.} The mechanochemical cycle of kinesins and the force versus extension curves (FECs) for the NL of red motor domain. There are three FECs; two for the unzippered NL (cyan) and one for the zippered NL (purple). Each stage of kinesin cycle is marked with the circles (yellow) on the FECs indexed from (i) to (viii). }
 \label{FEC_cycle}
 \end{center} 
\end{figure}

In contrast to the thermodynamic conditions imposed on macroscopic engines, 
molecular motors operate under isothermal and highly dissipative environments; thus an  thermodynamic cycle with two distinct isotherms as in the Carnot engine is not applicable for biological motors (Fig.~S1).     
Instead, conformational changes of internal structure can be linked to the thermodynamic cycle.  
As a simple example, the thermodynamics of the optomechanical coupling in polyazopeptide \cite{Hugel02Science} is best described in terms of two thermodynamically conjugate variables, force ($f$) and extension ($x$) (Fig.~\ref{FEC_cycle}a). 
Photon absorption/emission transforms the polymeric property of polyazopeptide, the persistence length ($l_p$) and contour length ($L_c$), via $cis\leftrightarrow trans$ conformational change. 
The two distinct force-extension curves with different $l_p$ and $L_c$, corresponding to the distinct conformational states of polyazopeptide, are analogous to the two isotherms in the Carnot cycle.  
The area enclosed by the optomechanical cycle  
is the maximal mechanical work that the polymer engine can extract from the photon energy when the polymer is stretched and relaxed quasi-statically. 
Biological machines adopt a similar strategy as the polyazopeptide by changing molecular conformations in response to the interactions with molecular fuels. 
In analogy to polyazopeptide whose mechanical property and conformational change are used to generate work, it is essential to identify relevant structural elements, responsible for internal mechanics, from its complicated architecture, in order to understand the work generating mechanism of each biological motor.

For kinesin, as suggested in the previous sections, the NL is one of the key structural elements responsible for both the stepping dynamics and head-head regulation.    
Along the biochemical cycle, the change in motor head affinity to MTs as well as the disorder-order transition of the NL alters the polymer property of NL. 
By assuming that NL, a 15 amino acid polypeptide segment, is well described using worm-like chain model (see, however, Ref \cite{kutys2010PLOSC}), we  propose a diagram in Fig.~\ref{FEC_cycle}b that depicts the magnitude of an internal tension ($f_{int}$) built on one of the NLs (NL of red motor head) versus its extension ($\delta x$) with its direction ($\delta x>0$ for forward and $\delta x<0$ for backward extension) for the full cycle. 
The diagram of thermodynamic cycle, built by tracking the state of NL along the biochemical cycle of kinesin, provides a integrated view of how the extension/release and conformational change of NL lead to generate work and regulate the head-head coordination.

In (i)$\rightarrow$(ii), the NL changes both the length and direction upon docking transition. In this step the residue 326 binds the $\beta 10$ to form a $\beta$ sheet, leaving only two residues at the C-terminal of the NL. 
As a result, the contour length of the NL ($L_c$) decreases from $L_c\approx 5.7$ nm to $\approx $1.0 nm. 
The $L_c\approx 1.0$ nm is maintained until Pi is released from the trailing (red) head at the state (v). 
Upon ADP released at (ii)$\rightarrow$(iii), the leading (blue) head strongly binds the MT.  
In (iii)$\rightarrow$(iv)$\rightarrow$(v), Pi release from the trailing head is followed by the order-disorder transition in the NL, which restores the contour length into $L_c=5.7$ nm. 
If (v)$\rightarrow$(vi) step were to occur quasi-statically, force mechanics of the NL would follow the line of the force-extension curve. Along (vi)$\rightarrow$(vii)$\rightarrow$(viii)$\rightarrow$(i), the value of $L_c$ does not change but because the binding affinity of the trailing (blue) head is altered, the $f_{int}$ built on the NL is expected to vary. 

The thermodynamic cycle of kinesin by focusing on the force extension property of one of the NLs (Fig.~\ref{FEC_cycle}b) provides a few interesting points.
First, the $f_{int}$ would be greatest when the two motor heads are simultaneously bound to the MT. By assuming $l_p\approx 0.6$ nm for the NL, the $f_{int}$ for the two-head bound (2HB) state (see the section discussing the nature of ATP-wait state in SI text) is expected to be $f^{2HB}_{int}\approx 11$ pN. 
It is of particular note that the estimated value of $f^{2HB}_{int}\approx 11$ pN should not be confused with the value of force ($f_{stall}\approx 6-7$ pN) externally applied through the coiled-coil stalk that can stall the kinesin motion \cite{Yildiz08Cell}.   
 While $f_{stall}$ interferes with 
 the step (i)$\rightarrow$(ii) or (v)$\rightarrow$(vi) by preventing the NL zipper process,  
 the $f_{int}^{2HB}$ on the NL is used to regulate the head-head coordination. 
 In addition, the diagram suggests that $f^{2HB}_{int}>f_{stall}$. 
Second, the diagram of thermodynamic cycle succinctly indicates the origin of symmetry breaking, which gives rise to the motor directionality. The finite area enclosed by force extension curves is formed along the process from (i) to (vi) while the process involving (vi)$\rightarrow$(vii)$\rightarrow$(viii)$\rightarrow$(i) creates no work. Among the series of different kinesin configurations, the work generating conformational change of NL occurs while the motor head is mostly in the trailing position ($\delta x>0$) and its configuration is sequentially altered from ATP state to ADP state (see the red head in Fig.~\ref{FEC_cycle}b). 
Third, during the process of conformational change in red head spanning the half-cycle, the partner (blue) head is mostly in NT free state, waiting for the enzymatic cycle of the red head to be completed. It is noteworthy that although the process (vi)$\rightarrow$(vii)$\rightarrow$(viii)$\rightarrow$(i) creates no work, during which the NL of the red head points the rearward direction ($\delta x<0$), the $f_{int}$ whose value changes between $f_{stall}$ and $f_{int}^{2HB}$ is used to inhibit the ATP binding, which recapitulates the origin of head-head coordination.
      
We used the thermodynamic cycle for the NL of kinesin (Fig.~\ref{FEC_cycle}b), similar to the optomechanical cycle of polyazopeptide (Fig.~\ref{FEC_cycle}a), to highlight the NL as a key element of work generation and regulation in kinesin dynamics.

\section*{Concluding Remarks.}
In this paper, using coarse-grained molecular simulations and theoretical considerations based on the theories of protein folding and polymer dynamics, we presented our perspective on how the kinesin, under significant thermal fluctuation, orchestrates its structure, conformational dynamics, and interaction with NT and MTs to constitute the characteristic biochemical cycle.     
We adopted into our coarse-grained model the hypothesis of structure-function relationship \cite{Bryngelson95Protein} from the study of protein function and its extension to protein-protein and protein-ligand interactions, and tried to address several key issues in the kinesin dynamics. 
Although a series of rate constants that constitute the enzymatic cycle may allow us to formally explain the behavior of kinesin motor as a function of external force or concentration of ATP \cite{astumian2010BJ}, the chemical specificity of the motor itself, encoded in the microscopic rate constant and affinity to the MT, is determined by the structural details at each stage of the cycle. 
It is impressive to see a number of recent experimental efforts to engineer the chemical specificity of kinesin by altering various structural elements, which includes changing the length of NL \cite{Hackney03Biochem,Yildiz08Cell,clancy2011NSMB}, mutating the amino-acid at the neck region \cite{endow2000Nature}, and removing the negatively charged C-terminal region of tubulin (E-hook) \cite{Hirokawa00PNAS,lakamper2005hook}. 
The ``mechanical" notions we tried to employ with our computational model in this paper such as stress, strain, deformation, tension, persistence length should be useful to clarify why the microscopic rate constant or affinity has that particular value and provide better idea to understand and engineer the dynamics of biological motors.

Lastly, the notion of motor efficiency is worth mentioning.  
Although the literature often discusses the unusually high thermodynamic efficiency of biological motors \cite{Oster00JBB,kinosita2000rotary}, 
the thermodynamic efficiency itself may not be such a relevant measure to understand the principles of biological motors in the cell where the amount of molecular fuels are buffered by the cellular metabolism. 
Depending on the functional goal of a biological motor, the optimization criteria may vary \cite{Derenyi99PRL}.
Analysis of the energy balance of kinesins \cite{Hackney05PNAS} indicates that at physiological condition, 60\% of the free energy stored in ATP is expended upon ATP binding to induce the conformational changes directly linked to the stepping dynamics; the remaining 40\% of free energy due to the sequence of processes followed by ATP binding is \emph{partitioned} into several microscopic steps that do not produce any work \cite{Hyeon09PCCP}. 
As discussed throughout the paper, the \emph{non-work producing steps}, depicted with purple arrows in Fig.~\ref{cycle_kinesin} or the processes corresponding to (vi)$\rightarrow$(vii)$\rightarrow$(viii)$\rightarrow$(i) in Fig.~\ref{FEC_cycle}b, are responsible for regulating the MT affinity of kinesin motor domain, the configurational state of NL, and the NT state of partner head through head-head coordination. Similar to the concept of \emph{kinetic proofreading} to increase the fidelity of biological processes \cite{Hopfield74PNAS}, the series of non-work producing steps, which otherwise appear to be aimless side reactions, are essential for kinesins to have the persistent motor function.   
To regulate the molecular configurations under excessive thermal noise, biological motors reserve substantial amount of free energy. Understanding the mechanism of allosteric regulation at the microscopic level would be one of the important topics to explore in the study of biological motors. 

\section*{Acknowledgments}
We are grateful to R. D. Astumian for illuminating discussion on the role of mechanical components of molecular motors in determining the chemical specificity, and to Paul Whitford for carefully reading this manuscript.  
This work was supported in part by the grants from the National Research Foundation of Korea (NRF)  (2010-0000602) (to C.H.) and by the Center for Theoretical
Biological Physics sponsored by the National Science Foundation
(NSF; Grant PHY-0822283 and  
MCB-1051438) (to J.N.O.).

%\bibliographystyle{pnas2010}
%\bibliography{mybib1}

\clearpage
\begin{table}[ht]
\centering
\begin{tabular}{|c||c|c|}
	\hline
  &  NL state \cite{ValeNature99}  &  Affinity to MTs  \cite{Cross00PTRSL}   \\
	\hline
$K\cdot\phi$   & disordered (off) & strong ($K_d\sim nM$) \\
$K\cdot$ATP    & ordered (on) & strong \\
$K\cdot$ADP$\cdot$Pi   & ordered (on) & strong \\
$K\cdot$ADP   & disordered (off) & weak ($K_d\sim \mu M$) \\
	\hline
\end{tabular}
\caption{The NT state dependent configuration of a kinesin monomer}
\label{Table}
\end{table}

\clearpage

% Figures, one per page (fig_1.eps and fig_1.pdf files must be present
% in the document directory)

% closing statement, nothing below matters
\clearpage 

\noindent{\bf\Large SUPPORTING INFORMATION}

\section*{Operational conditions for biological machines}
In the macroworld, the length, time, and energy scales of an object are well-separated from its surrounding, thus conventional Newtonian mechanics can determine the trajectory of the object deterministically. 
Such separation of scales, however, does not hold as the object size is reduced to nanoscales where  the effect of thermal noise from the surrounding is substantial. 
While the thermal noise imparts incessant fluctuation to the molecule and hinders precise measurement, 
the thermal noise is an essential component to trigger the conformational changes of biological motors. 
Below we illuminate the general design principles of molecular machines by contrasting the operational condition of nanoscopic motors with that of macroscopic counterparts.\\
 
\subsubsection*{Energy scales}
For biomolecules whose sizes are on the nanometer scale, the effect of thermal noise ($k_BT\approx 4\times 10^{-21}J=4pN\cdot nm$) on the molecule is no longer negligible because the energy scale of individual non-covalent bond interactions ($\sim\mathcal{O}(1)$ $k_BT$) is of the same order.  
Dominance of entropy is ubiquitous in soft matters such as polymer, colloids, bubbles and biomolecules. 
Due to free energy barriers that are created by incomplete cancellation between enthalpy and entropy \cite{OnuchicCOSB04}, conformational transitions of biopolymers occur cooperatively.  
If the kinetic barrier associated with a transition is too large to overcome using thermal fluctuations, free energy borrowed from the chemical potential of molecular fuel can make the conformational change biologically accessible.    
Actively adopting the environmental noise to induce functional motion is one of the key design principles unique to biological motors. 
In contrast, for macroscopic motors that operate through tight couplings among a multitude of rigid bodies, the thermal fluctuations are rather a nuisance to achieve better accuracy and precision \cite{oosawa2000loose}. 
An integral part of this review lies in understanding the \emph{soft} mechanics of biological motors.

\subsubsection*{Thermodynamics} 
Carnot engines that exemplifies the thermodynamics in the macroworld extract mechanical work ($W$) by accepting heat $Q_h$ from a hot reservoir at temperature $T_h$ and discarding the remaining heat $Q_c$ into a cold reservoir at $T_c$ (see Fig. \ref{macromachine}a for gasoline engine as an example of practical heat engine cycle). 
The maximal work, which amounts to the heat transfer between the two reservoirs $W_{max}=Q_h-Q_c$, is extracted when an engine is operated \emph{quasi-statically}without dissipating heat. 
%In terms of pressure ($p$) and volume ($V$) as two probe variables, the maximum work corresponds to the area enclosed by quasi-static $pV$ curves.  
Real heat engines, however, do not adopt the quasi-static operation for all practical purposes; some amount of heat dissipation is inevitable. 
As a result, the extracted work is always smaller than the heat transfer ($W\leq Q_h-Q_c$), thus the thermodynamic efficiency for the engine is bounded by $\eta_{Carnot} = (Q_h-Q_c)/Q_h=1-T_c/T_h$, where $Q_h/Q_c=T_h/T_c$. 
To increase the efficiency, macroscopic machines are designed to minimize heat dissipation by lubricating the parts of the machine that are prone to produce friction.    
%Since the heat dissipation is proportional to the surface area of an engine, macroscopic heat engines are designed in bigger size for the better efficiency. 

In contrast to the thermodynamic conditions imposed on macroscopic engines, 
molecular motors operate under isothermal and highly dissipative environments; thus an  thermodynamic cycle with two distinct isotherms as in the Carnot engine is not applicable for biological motors.    
Instead, conformational changes of internal structure can be linked to the motor cycle (see the main text for the discussion using polyazopeptide and kinesin).  
%As a simple example, the thermodynamics of the optomechanical coupling in polyazopeptide \cite{Hugel02Science} is best described in terms of two thermodynamically conjugate variables, force ($f$) and extension ($x$) (Fig.\ref{machine-cycle}b). 
%Photon absorption/emission alters the polymeric property of polyazopeptide, the persistence length ($l_p$) and contour length ($L_c$), via $cis\leftrightarrow trans$ conformational change. 
%The two distinct force-extension curves with different $l_p$ and $L_c$, corresponding to the distinct conformational states of polyazopeptide, is analogous to the two isotherms in the Carnot cycle.  
%The area enclosed by the optomechanical cycle  
%is the maximal mechanical work that the biological engine can extract from the photon energy. 
%Biological machines adopt a similar strategy as the polyazopeptide by changing molecular conformations in response to the interactions with molecular fuels such as ATP, Ca$^{2+}$ \cite{lee2010Science}, and [H$^+$] gradient across ion channels. 
%To understand the detailed mechanism of each biological motor, it is essential to identify relevant structural elements hidden in its complicated architecture. 
%For kinesins, it will be shown that the neck-linker, a 15 amino acid spacer between motor head and neck-helix, plays vital roles not only in generating work but also in regulating the persistent  mechanochemical cycle. 

\subsubsection*{Low Reynolds number environment}
Reynolds number, defined as $Re=\rho v a/\eta$, is equivalent to the ratio of inertial force ($F_{inertial}\propto\rho a^2v^2$) to viscous force ($F_{viscous}\propto\eta a v$), where $\rho$ and $\eta$ are the density and viscosity of media, and $a$ and $v$ are the size and speed of an object in motion \cite{Purcell77AJP}. 
Significant variation of $Re$ values implies that there is fundamental difference in the dynamics.   
The Reynolds numbers associated with motions at both atomic and macroscopic scales satisfy $Re\gg 1$.  
A macroscopic object ($a\sim cm$) moving $v\sim cm/s$ through water with the kinematic viscosity $\eta/\rho\approx 10^{-2}$ $cm^2/sec$ will have $Re\sim 10^2\gg 1$;   
At atomic scales, the surrounding media looks discrete, so the dynamics such as vibrational motion of hydrogen bonds occurs essentially in vacuum ($\eta/\rho\approx 0$), thus $Re\gg 1$ and inertial forces are dominant.     
In contrast, for a typical protein, whose size is $a\sim nm$, the time scales associated with domain motions, such as looping dynamics and beta sheet formation is $\lesssim (0.1-1)$ $\mu sec$ \cite{Thirum05Biochem}; thus the typical velocity of motion is $v\approx 0.5$ $cm/sec$, resulting in $Re\approx (10^{-5}-10^{-6})\ll 1$.     
The estimated value of $Re$ indicates that even for a large biomolecule like the ribosome ($a\sim 20$ $nm$) the molecular motions occur in a very low $Re$ regime. 
% ($F_{inertial}\ll F_{viscous}$). 
In the $Re\ll 1$ regime, reciprocal motions employed by a swimmer in macroworld do not gain propulsion from the fluid surrounding the molecule. 
Molecular motors, therefore, have to adopt an entirely different strategy from the one in the macroworld, to break time reversal and translational symmetry \cite{Purcell77AJP}.     
\\

% affects the accessible conformational space sampled via molecular fluctuation. 
%This feature of biological motors is different from that of macroscopic engines in that topology of macroscopic engine is kept constant or at least macroscopic engines go through elastic deformation along the engine cycle.  
%The different molecular topology at each stage of biochemical cycle implies that accessible conformational space sampled via molecular fluctuation changes as reaction proceeds.  

%Fig.** demonstrates the contact map of KIF1A, a single headed kinesin that moves along the microtubule through biased diffusion.  
%Every reaction of ATP hydrolysis produces a nonvanishing area enclosed by multiple pieces of force-extension curves.  

The above mentioned basic constraints enable us to draw a few general conclusions about molecular motors:  
(i) Due to constant exposure to noisy environment and structural flexibility, 
%the conformational change in response to environmental stresses is stochastic.  
the deterministic description using the rigid body motion of macroscopic objects is no longer valid to describe the probabilistic nature of dynamics of biological systems. 
(ii) 
Unlike macroscopic heat engines that extract work out of heat reservoirs, 
molecular motors transduce chemical energy into mechanical work by exploiting the changes of molecular topology  \cite{Hugel02Science}.   
Small and large structural adaptations are the vital components of free energy transduction.   
(iii) %Typical estimates of thermodynamic efficiency ($=W/\Delta \mu_{ATP}$) by considering the ATP hydrolysis free energy ($\Delta \mu_{ATP}$) per enzymatic cycle and \emph{quasi-static} work of the motor estimated under stall condition ($W=f_{stall}\times\delta x$) are $\sim 60$ \% for kinesin \cite{Cross05Nature,Nishiyama02NCB}, $\sim 90-100$ \% for myosin \cite{mehta99Nature,rock2001PNAS}, and $\sim 100$ \% for F$_1$-rotary motor \cite{Noji97Nature,kinosita2000rotary}, which are significantly larger than a value one can obtain in any macroscopic machine. 
For biological motors bombarded by thermal noise, %however, 
the distinction between the system of interest and surroundings is not obvious since system-bath interactions are non-negligible. 
Furthermore, unlike macroscopic heat engines, 
the heat dissipated due to friction can also be replenished by the thermal energy from the solvent, which is a qualitative description of fluctuation-dissipation theorem.  
Instead of minimizing the heat dissipation, molecular machines actively utilize molecular fluctuations due to thermal noise.
(iv) To produce uni-directional transport or rotational dynamics under low Reynolds number conditions, isotropy in the diffusive dynamics has to be broken.  

%Just like many different kinds of man-made machines, one can easily find a multitude of motors in the cell. 
Besides the general principles of molecular motors, understanding individual motors requires more specific knowledge on their architecture and related dynamics \cite{Alberts1992Cell}.
To this end, we explore the details of design principles in biological motors by focusing on one of the best studied molecular motors, kinesin-1. \\

\section*{Nonequilibrium steady state thermodynamics} 
In discussing the action of molecular motors, it should be remembered that each state of the molecular motor along a biochemical cycle is in nonequilibrium steady state (NESS); in principle, one should not use the notion of equilibrium thermodynamics. 
Nevertheless, an extended form of the thermodynamics for NESS has long been suggested and recently revisited \cite{Hill1989,QianBook}. In the NESS formalism, the second law of steady state thermodynamics reads $T\Delta S\ge Q_{ex}$ or $\langle W_{ex}\rangle\ge \Delta F$ where $\Delta S$ is the difference of Shannon entropy between the two states and $Q_{ex}$ (excess heat) is the total heat ($Q_{tot}$) subtracted by the housekeeping heat ($Q_{hk}$) generated in an infinitely slow process, where $Q_{hk}$ is the heat required to maintain constant non-zero flux between the two steady states \cite{Hatano01PRL}. 
According to the NESS thermodynamics of open systems, the concentrations of four NT states, held constant as a source and sinks of the system, provide a constant chemical potential difference that drives the biochemical cycle; one can define an effective free energy difference between two steady states by using the ratio of the forward and backward rate constants measured at near-equilibrium condition \cite{Hill1989,QianBook,Hyeon09Bookchap,Hyeon09PCCP}.  

\section*{ATP-Wait state}
%Depending on the concentration of ATP, the rate determining step of biochemical cycle is altered.
At high [ATP]($\approx$ 1 mM) %, which is the physiological ATP concentration, 
when the leading head is bombarded by ATP molecules every 0.5 ms (bi-molecular rate constant for ATP binding is $k_{ATP}^o=2.0\pm 0.8\mu M^{-1}s^{-1}$), %. In this case, 
the rate limiting step for the kinesin cycle is the hydrolysis or Pi-release from the trailing head ($\approx 10$ ms).  
In contrast, at [ATP]$\approx$ $1\mu M$, the leading head is inaccessible to ATP for 0.5 sec ($\gg 10$ ms) during which all the processes including ATP hydrolysis, Pi release take place, thus the ATP binding becomes the rate limiting step.
Therefore, among the series of kinesin configurations, 
the ADP-$\phi$ state becomes the rate limiting configuration at low [ATP]. 
It is worth reviewing the series of recent studies, led by several experimental groups, to address an intriguing question as to whether kinesins in ADP-$\phi$ state wait for ATP binding in one head (1HB) or two head bound (2HB) form \cite{Cross07Science,Mori07Nature,Asenjo09PNAS,Guydosh09Nature}. 

By analyzing the concentration of $\alpha\beta$-tubulin and kinesin-1 at various nucleotide conditions Alonso \emph{et al.} showed that stoichiometry of kinesin-1 to tubulin is 1:1 under ADP only condition but this changes to 1:2 in the presence of AMP-PNP (ATP-analog)  \cite{Cross07Science}. 
The 1:2 stoichiometry is consistent with the Table~\ref{Table} since both NT-free and AMP-PNP states strongly bind the MT. 
The 1:1 stoichiometry under ADP only condition results from a kinesin conformation in which one head is in NT-free state that binds MT strongly while the other head is still in ADP state.  
Based on the cryo-EM map,  
Alonso \emph{et al.} proposed a structure of ATP-wait state in which the tethered head parks on top of its partner head, blocking its MT-binding motifs, and further inferred that the ATP binding to the MT-bound NT-free kinesin head unblocks the MT binding motifs of the tethered head, enabling the kinesin-tubulin interaction.
 
Alonso \emph{et al.}'s proposal for ATP-wait conformation \cite{Cross07Science} was examined by the more quantitative analysis using sm-FRET, which directly probed the ``distance" between the two kinesin heads parked on MTs. 
In reference to the sm-FRET distributions for 2HB and 1HB structures, the latter of which were prepared using mutant heterodimeric kinesin, Mori \emph{et al.} concluded that kinesin waits ATP in 2 HB state at high [ATP]($\approx 1mM$) and in 1 HB state at low [ATP]($\approx 2\mu M$). 
At low [ATP], FRET time trace indicates that a kinesin step has a short-lived 2HB state, which then undergoes a transition to a long-lived 1HB state. 
Also, at low [ATP] the detached head is on average located rearward relative to the NT free MT-bound head \cite{Mori07Nature}. 

The picture of ATP-wait state by Mori \emph{et al.} \cite{Mori07Nature} was further elaborated by Asenjo \emph{et al.} who probed the mobility of tethered kinesin head by using fluorescence polarization microscopy (FPM) \cite{Asenjo09PNAS}. 
The FPM showed that the unbound head is highly mobile. 
Especially, the high mobility of detached head at low [ATP] rules out the parked configuration of the tethered  head proposed by Alonso \emph{et al}. % based on the cryo-EM map.
In fact, the PMF calculated for the head tethered to a disordered NL (Fig.~2b left) succinctly demonstrates the space explored by the tethered head in 1HB state \cite{Hyeon07PNAS2}.
The recognition of the next binding site is promoted only after the NL docking. 
The NL zippered state of the MT-bound head restricts the search space of the tethered head and facilitate the tethered head to locate the next MT-binding site. 

Most recently, Guydosh \emph{et al.} conducted an optical tweezers experiment to provide a direct probe of the ATP-wait stage at low ATP concentration (=$2 \mu M$) \cite{Guydosh09Nature}. Instead to the stalk region, they tethered a micro-bead to one of the two kinesin heads and probed its dynamics. In response to the external load that rapidly alternate between hindering ($-1.7$ pN) and assisting ($+1.7$ pN) phases with 14 $ms$ time interval, the kinesin head showed signals swinging back and forth, which indicate that the kinesin is in 1HB state. Such a swinging motion would have not been possible if the kinesin had been in 2HB state. Under the assisting load ($+1.7$ pN) the tethered head showed overshoot step that reaches $\sim 23$ nm and recovery step ($\sim -7$ nm) upon ATP binding to the MT-bound head. The decrease of overshoot dwell with increasing ATP concentrations suggests that the recovery step is induced by ATP binding. Based on the swinging motion that freely occurs with no signature of disruption even under $\pm 1.7$ pN load, Guydosh \emph{et al.} concluded that the ATP wait stage of kinesin is 1 HB state. The possibility of 2 HB state was essentially precluded based on the estimate that the binding affinity of weakly bound ADP state is even less than thermal energy $k_BT$ (If the location of transition state is assumed to be 2 nm, unbinding energy is 1.7 pN $\times$ 2 nm $\approx$ 3.4 pN$\cdot$nm$<$ 4.14 pN$\cdot$nm). The conclusion against 2HB state in Guydosh \emph{et al}'s study contradicts to Mori \emph{et al.} and Asenjo \emph{et al.}'s studies \cite{Mori07Nature, Asenjo09PNAS}. However, the time interval of alternating load ($\sim 14$ ms) used in the experiment is still longer than the average dwell time of kinesin at saturating ATP condition under which 2HB state is shown to be dominant in ADP-$\phi$ state in Refs. \cite{Mori07Nature, Guydosh09Nature}. A signature of disruption missing under 1.7 pN load may be due to the relatively long time interval during which the spontaneous dissociation of ADP containing tethered head can occur from the trailing head position in MTs.

\renewcommand{\thefigure}{S1}
\begin{figure}
\begin{center}
\includegraphics[width=5.5in]{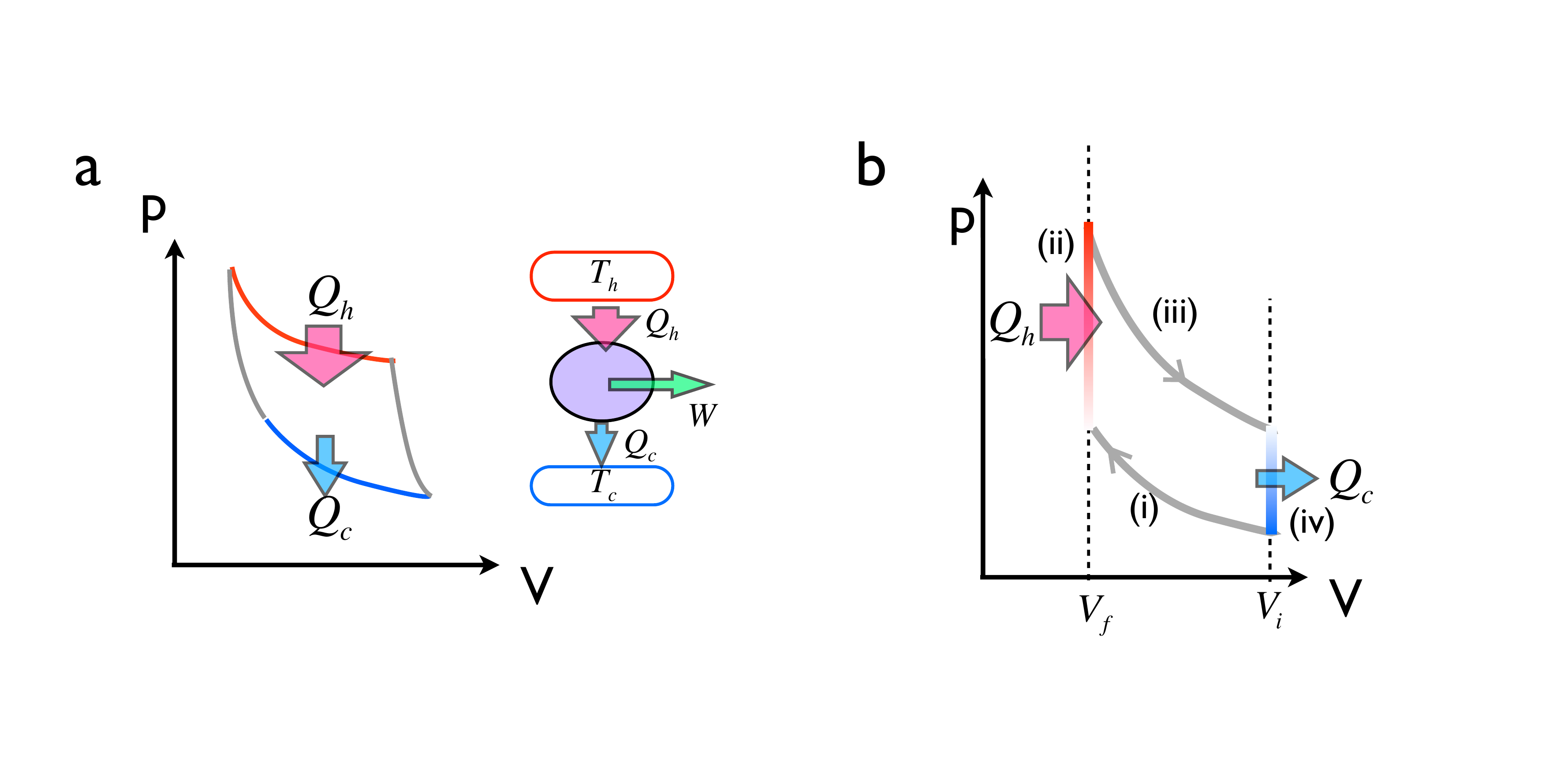}
 \caption{Macroscopic heat engines. 
%{\bf a.} A cartoon of an idealized macroscopic machine and a nanoscopic kinesin motor. Using its rigid body architecture, the macroscopic machine deterministically converts the energy from external heat source to the work with irreversible dissipations. Whereas, the kinesin is characterized by a soft architecture that is versatile to adapt its conformation, taking advantage of thermal noise as well as ATP binding to make conformational changes and produce work with high efficiency.  
{\bf a.} Thermodynamic cycle of Carnot engine.
%, whose maximal thermodynamic efficiency is given by $\eta_{Carnot}=1-T_c/T_h$, . 
%that extracts mechanical work ($W$) by absorbing $Q_h$ and removing $Q_c$ from two heat sources at temperatures $T_h$ and $T_c$. Without heat dissipation in the cycle maximal thermodynamic efficiency of this cycle is given by the ratio of temperatures at two heat sources $\eta_{Carnot}=1-T_c/T_h$.  
%In reality, however, the thermodynamic setup is more complicated. 
{\bf b.} The thermodynamic cycle of idealized gasoline engine (Otto cycle), which consists of (i) compression of vapor (adiabatic compression) $\rightarrow$ (ii) combustion of the gasoline at the chamber with constant volume (isochoric heating) $\rightarrow$ (iii) power stroke (adiabatic expansion) $\rightarrow$ (iv) isochoric cooling. 
At the step (ii) heat is absorbed to the engine but unlike Carnot cycle the temperature involving this step is not constant. Therefore, the absorbed heat should be integrated over the varying temperatures. 
By assuming an ideal gas for the gasoline vapor, the maximum thermodynamic efficiency of Otto cycle is obtained with the ratio of volume before and after the compression, and two heat capacities ($c_p=(dQ/dT)_p$, $c_v=(dQ/dT)_v$), $\eta_{Otto}=1-(V_{f}/V_{i})^{\frac{c_p-c_v}{c_v}}$. 
In practice, the engine needs to be supplied with fresh gasoline vapor and the compression cannot be made by the single engine alone. Thus, to realize the steps (i) and (iv) the piston of engine is operated in concert with others. 
%{\bf b.} A molecular motor devised by using optomechanical cycle of polyazopeptide that undergoes $cis\leftrightarrow trans$ transition upon photon absorption/emission. The figure is adapted from \cite{Hugel02Science}.    
}
 \label{macromachine} 
\end{center}
\end{figure}

\renewcommand{\thefigure}{S2}
\begin{figure}
\begin{center}
\includegraphics[width=5.0in]{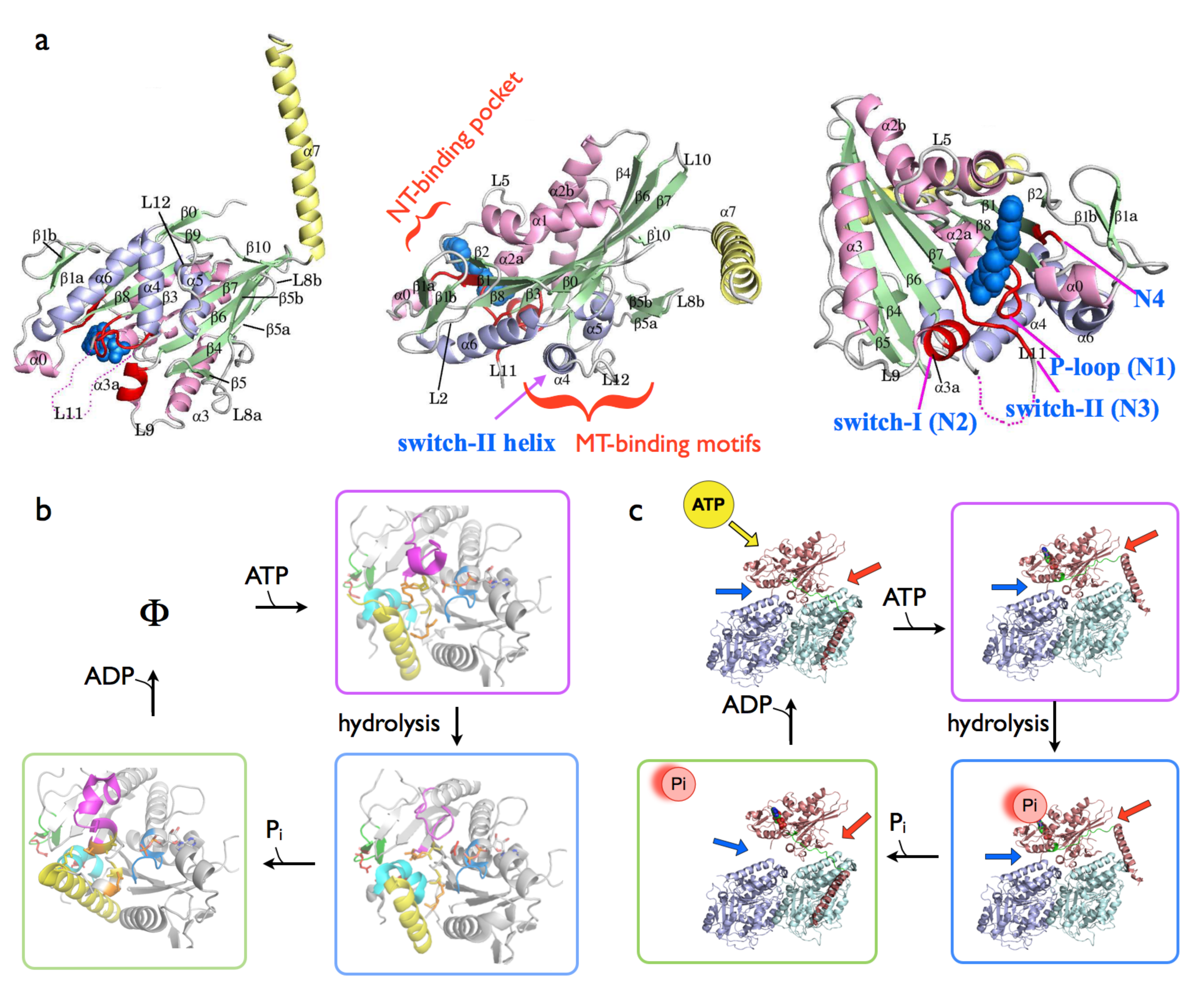}
 \caption{Kinesin structure and its conformational changes with various NT states.  
{\bf a}. Details of kinesin structure. 
The kinesin monomer is structurally divided into three regions: the head domain (or motor domain) (residues 2-323), the neck-linker (NL) (residues 324-338: $\beta$9, $\beta$10), and the neck-helix (residue 339-: $\alpha$7).  
With respect to the central $\beta$-sheets ($\beta1$, $\beta8$, $\beta3$, $\beta7$, $\beta6$, $\beta4$) the head domain is divided into two sides; one side, made of $\alpha4$, $\alpha5$, $\alpha6$, $L8$, $L11$, and $L12$, is used to bind MTs, and the other side has NT-binding pocket surrounded by the structural motifs consisting of the P-loop (N1) (86-93), switch-I (N2) (199-204), switch-II (N3) (232-237), and N4. 
Structural changes of kinesin along the biochemical cycle of ATP binding, hydrolysis, Pi and ADP releases are shown in {\bf b} and {\bf c}. {\bf b}. The configurations of structural motifs surrounding the nucleotide binding pocket in AMPPNP (ATP analog) state (PDB id : 1i6i) (enclosed in magenta box), in ADP-AlF$_x$ (ADP$\cdot$Pi analog) state (PDB id : 1vfx) (blue box), and in ADP-V$_i$ (ADP analog) state (green box) \cite{HirokawaSCI04}. 
%{\bf c}. KIF1A structures bound to the tubulin dimer in AMPPNP state (red, PDB id : 2HXF) and in ADP state (yellow, PDB id : 2HXH). A tilt of $\alpha4$ helix resulting from the conformational switchings of other motif leads to a 20$^o$ rotation of the whole molecule \cite{Kikkawa06EMBOJ}.
The orientation of $\alpha4$ (yellow helix) changes upon Pi release. 
{\bf c}. Schematics showing the changes of NL configuration and kinesin-MT interaction indicated with red and blue arrows, respectively. 
}
 \label{kinesin_struct} 
\end{center}
\end{figure}

\renewcommand{\thefigure}{S3}
\begin{figure}
\begin{center}
\includegraphics[width=5.5in]{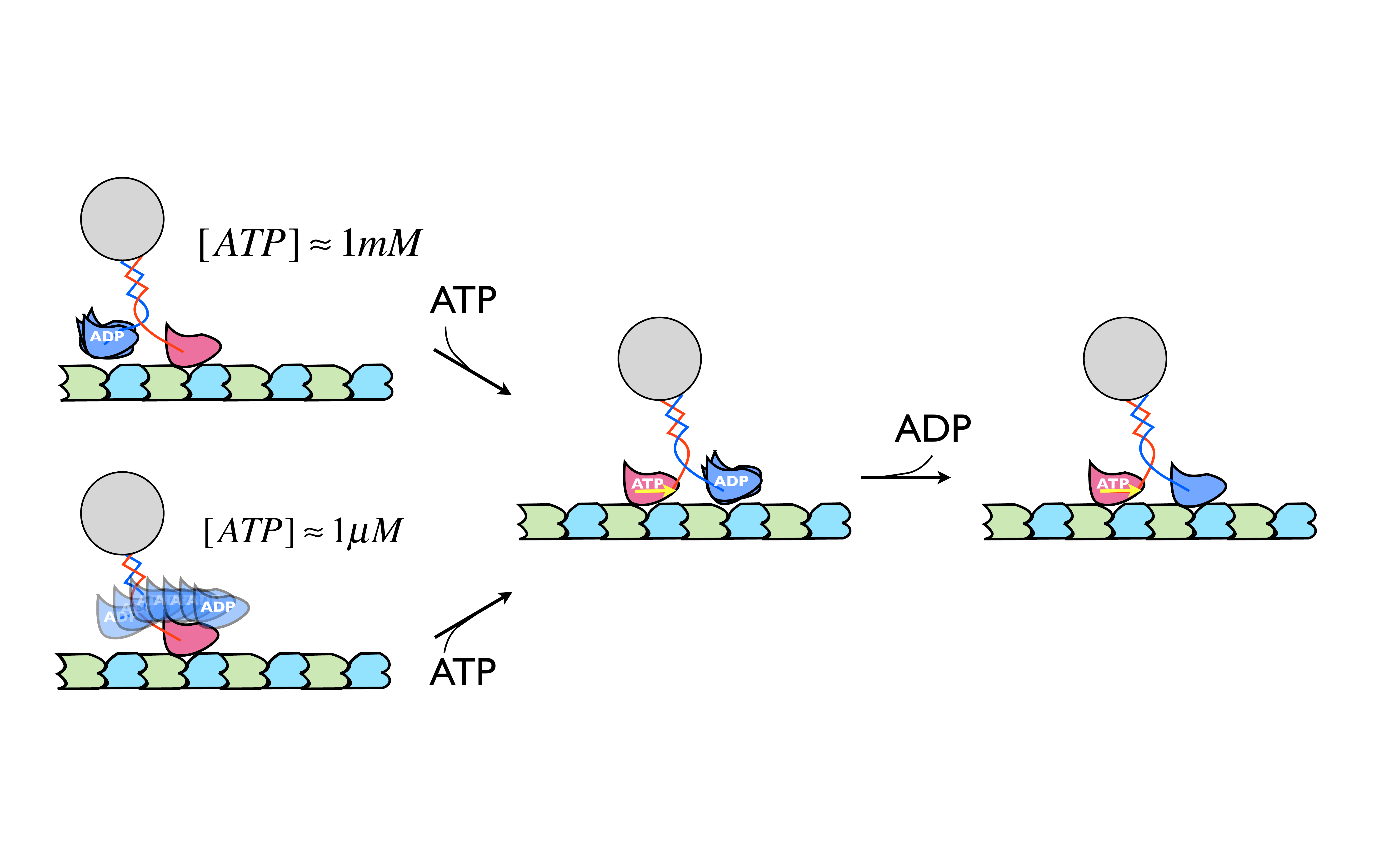}
 \caption{ATP-wait state at two different ATP concentrations. At high ([ATP]=1mM) and low ([ATP]=1 $\mu$M) ATP concentrations, kinesins wait for ATP in 2HB and 1HB states, respectively. 
At 1HB state, the cartoon illustrate a substantial fluctuation in the tethered head \cite{Asenjo09PNAS}.  
}\label{ATP-wait}
 \end{center}
\end{figure}

\end{document}